\documentclass{article}
\usepackage{ijcai25}

\usepackage{times}
\usepackage{soul}
\usepackage{url}

\usepackage[hidelinks]{hyperref}
\usepackage[utf8]{inputenc}
\usepackage[small]{caption}
\usepackage{graphicx}
\usepackage{amsmath}
\usepackage{amsthm}
\usepackage{booktabs}
\usepackage{algorithm}
\usepackage{algorithmic}
\usepackage[switch]{lineno}

\usepackage{subfig}
\usepackage{pifont}
\usepackage{comment}
\usepackage{booktabs}
\usepackage{tabularray}
\usepackage{xcolor}
\usepackage{bm}
\title{\method\ -- A Question-Bank-Based Approach to Fine-Grained Legal
Knowledge Retrieval for the General Public}

\author{
Mingruo Yuan\and
Ben Kao\footnote{Corresponding author: kao@cs.hku.hk}\and
Tien-Hsuan Wu\\
\affiliations
The University of Hong Kong\\
\emails
\{mryuan, kao, thwu\}@cs.hku.hk\\
}

\newcommand{\kevin}[1]{{\textbf{\textcolor{orange}{[Kevin: #1]}}}}

\newcommand{\edit}[1]{} % hide the markups

\newcommand{\method}{QBR}
\newcommand{\simplebullet}{\newline\noindent$\bullet$}
\newcommand{\miniheading}[1]{\noindent\textbf{[#1]}}
\newcommand{\qspair}[2]{(#1, #2)}

\newcommand{\qds}{(q, d_q, s_q)}
\newcommand{\qb}{\mathit{QB}}
\newcommand{\qbh}{\mathit{QB}_H}
\newcommand{\qbm}{\mathit{QB}_M}
\newcommand{\qd}[1]{Q_{#1}}
\newcommand{\sd}[1]{S_{#1}}
\newcommand{\simut}[3]{\mathit{Sim}_{#1}(#2, #3)}
\newcommand{\posex}[1]{E^+(#1)}
\newcommand{\negex}[1]{E^-(#1)}
\newcommand{\gptui}[1]{\widehat{u_{#1}}}
\newcommand{\mrrd}{\mathit{MRR}_d}
\newcommand{\mrrs}{\mathit{MRR}_s}
\newcommand{\acc}{\mathit{acc}}
\newcommand{\qbrnogpt}{\text{QBR}_{\neg\text{GPT}}}
\DeclareMathOperator{\argmax}{arg\,max}

\begin{document}

\maketitle

\begin{abstract}
% \kevin{\textit{Domain specific} or \textit{legal}}
Retrieval of legal knowledge %(such as legal and medical) 
by the general public is a challenging problem due to the technicality of the professional knowledge and the lack of fundamental understanding by laypersons on the subject. Traditional information retrieval techniques assume that users are capable of formulating succinct and precise queries for effective document retrieval. In practice, however, the wide gap between the highly technical contents and untrained users makes legal knowledge retrieval very difficult.
We propose a methodology, called QBR \footnote{\label{fn:appendix}\url{https://github.com/mingruo-yuan/QBR}}, which employs a Questions Bank (QB) as an effective medium for bridging the knowledge gap.
We show how the QB is used to derive training samples to enhance the embedding of knowledge units within documents, which leads to
effective fine-grained knowledge retrieval.
%Specifically, with QBR, knowledge retrieval involves two steps. First, given a user input $u$ that verbally describes a user's interest or concerns, QBR consults the QB and display a few
%semantically relevant questions in the QB to the user. Then, the user selects those questions presented that best express his/her search interests and QBR would then retrieve the relevant documents and identify specific sections in them that directly answer those questions. 
We discuss and evaluate through experiments various advantages of QBR over traditional methods. These include more accurate, efficient, and explainable document retrieval, better comprehension of retrieval results, and highly effective fine-grained knowledge retrieval. 
We also present some case studies and show that QBR achieves social impact by assisting citizens to resolve everyday legal concerns. 
\end{abstract}


\section{Introduction}
Law is inextricably linked to daily life.
Day-to-day activities are subject to legal regulations.
Whether we are shopping, at work, driving, or posting on social media, legal considerations are always at play. 
It is thus important for individuals to understand the law to protect their rights and benefits.
Yet, legal knowledge is extensive and technical.
It is impractical for one to master all aspects of law.
As such, we often rely on legal professionals for help when facing legal issues.

There are online platforms that provide support for individuals on legal issues, such as ABA Free Legal Answers\footnote{\url{https://abafreelegalanswers.org/}} in the US, LawWorks\footnote{\url{https://www.lawworks.org.uk/}} in the UK, and Justice Connect\footnote{\url{https://justiceconnect.org.au/}} in Australia.
Questions posted on those platforms are answered by pro bono lawyers for free.
There is a significant rise in the demand for these legal question-answering services.
\citeauthor{patino}~\shortcite{patino} report that legal problems are ubiquitous.
Their survey shows that approximately half of the people interviewed had experienced legal problems within two years prior to the interviews.
In addition, it is reported in \cite{aba:report} that the number of legal questions responded to in each year had increased from 4,193 in 2016 to 71,640 in 2023.
This increase suggests that more people are turning to online help.
However, professional, licensed pro bono lawyers are a scarce resource, which cannot meet 
the ever-increasing demand for online legal help.
In this work, we present \method, an AI-assisted approach that helps users retrieve relevant legal knowledge that addresses a user's 
legal situation by
leveraging a collection of legal educational articles and a question bank.
We show that our platform can serve a large community in providing expert legal answers,
thus effectively addressing the high demand for online legal question answering.

\begin{comment}
Providing legal aid is essential to the society as it promotes access to justice~\cite{teremetskyi2021access}. %,sommerlad2004some}.
%Citizens should not be denied of legal actions to protect their rights simply because they have no means or have no legal knowledge.
Legal aid, such as legal clinics that answer common legal questions and pro bono lawyers that represent clients in the court, 
ensures that there is a fair access to justice among citizens.
The benefits of access to justice are multifaceted~\cite{weston};
It reduces the pain and suffering of individuals who are facing legal problems,
as professional opinions would ease their burden and help resolve their problems faster.
In some cases, individuals may choose alternative pathways to save the legal costs and resolve the legal problem faster, if they have access to such legal knowledge.
For the society, the legal aid contributes to the values of equitable justice system as it provides support to vulnerable and marginalized groups.
In marginalized groups, the social contract (authority of the government over individuals) is often the weakest~\cite{ezer2018legal}.
A fair access to justice allows the marginalized group assistance claim their rights, which heals their relationship with the government.
To support access to justice, our \method\ provides free answers to legal questions and accesses to legal knowledge.
\end{comment}

Nowadays, legal information such as court judgments and legislation is available online in many countries.
%It is beneficial for the public to have the legal information available online; 
However, their online availability does not translate directly into effective public access to legal knowledge.
It remains challenging for ordinary people without a legal background to learn legal knowledge for two reasons. 
Firstly, the information available online consists mostly of primary legal sources, such as cases and statutes.
These documents are written in formal legal language that is generally hard to comprehend by the public. 
The incomprehensibility of professional documents has been reported in various studies \cite{examining_medical_reading,covid_1,covid_2}. 
For example, \cite{law_reading} analyzes the readability of 201 legislations and related policy documents in the European Union (EU). It is found that a PhD-level education is required to comprehend certain laws and policy documents.
Secondly, the public may not know the legal principles that are applicable to the legal situation they face. 
With large numbers of documents, it is difficult for a user to locate the correct legal sources 
in search of a solution to his/her legal problem. 
Information retrieval (IR) systems that can bridge such a knowledge gap are therefore essential.

\begin{figure*}
\centering
\subfloat[Traditional Information Retrieval Approach]{ 
	\label{fig:tradition_IR}\includegraphics[width=0.4\linewidth, height=2.1cm]{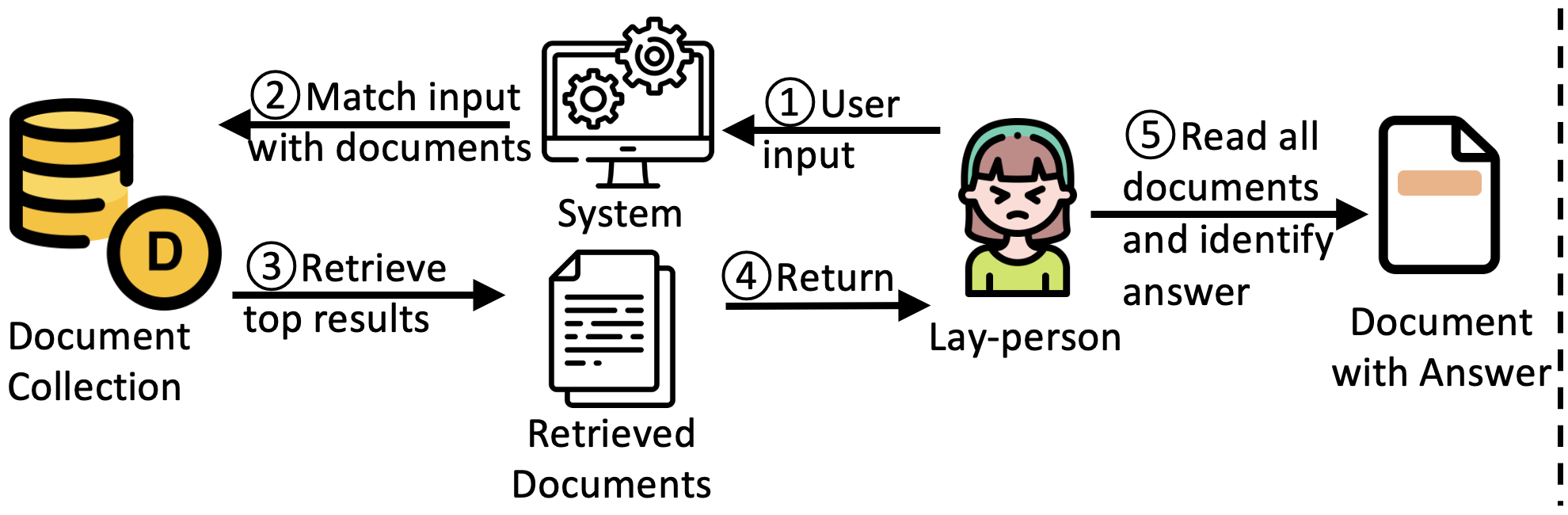}}
\subfloat[QBR Approach]{
	\label{fig:QBR}\includegraphics[width=0.45\linewidth,height=2.1cm]{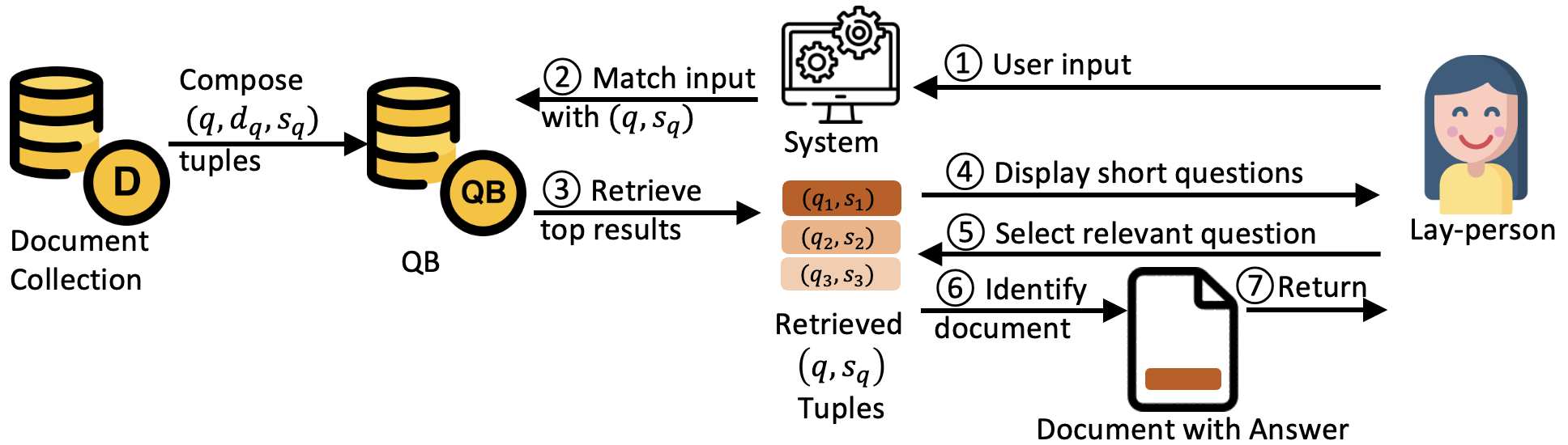}}
%\caption{Bridge the knowledge gap between the formal domain-specific knowledge and the non-expert general public.}
\caption{Traditional IR vs QBR}
\label{fig:task}
\end{figure*}

In this paper we propose a question-bank-based retrieval (\method) methodology for bridging the knowledge gap.
We start with a corpus of legal educational documents. These documents are written by lawyers to explain the various legal
concepts in layperson's terms. 
Based on the document contents, we generate model questions and answers, which  are collected in a {\it Question Bank} (QB).
\method\ answers a user's legal question through retrieving model questions and answers from the QB.
%an online legal education website that contains thousands of web pages written by lawyers to explain legal concepts to the general public.
In traditional IR, a typical search engine returns a shortlist of documents given a user input query. The user is expected to read the documents in the search result, understand each of them, pick the best matching one, and then extract the most relevant knowledge units (sentences or paragraphs within the documents) that provide an answer to the enquiry.
This workflow is illustrated in Figure~\ref{fig:tradition_IR}.
Due to the knowledge gap between the professional documents and novice users, it is generally a daunting task for a user to perform the post-search knowledge filtering and extraction step (Step \ding{176} in Figure~\ref{fig:tradition_IR}).

Our QBR approach (illustrated in Figure~\ref{fig:QBR}) aims at providing a more comprehensible search results to the user to greatly simplify this step.
The idea is to help a user (who may not have any legal training) express a legal question by selecting model questions from the QB.
Instead of returning a list of documents, QBR returns a list of {\it question-answer pairs} $\qspair{q}{s_q}$'s.
Specifically, for each $\qspair{q}{s_q}$ pair returned,
$q$ is a question from the QB and $s_q$ is an {\it answer scope}, which is 
a logical knowledge unit (typically consists of one or few paragraphs) extracted from a document $d_q$ such that $s_q$ answers the question $q$.
To filter the search results, a user needs only read the very short questions $q$'s returned to locate relevant ones. 
This is much simpler than reading the full contents of shortlisted documents under traditional search engines (Figure~\ref{fig:tradition_IR}). 
Moreover, the question $q$ effectively {\it explains} why the answer $s_q$ (and its originating document $d_q$) is relevant to the user's
enquiry; that $s_q$ (and thus $d_q$) answers the question $q$, which the system infers as a more proper expression of the user's enquiry. 
Table~\ref{tab:example} shows an example user input and one returned $\qspair{q}{s_q}$ pair.
In this example the answer scope $s_q$, which is highly relevant to the user's legal situation, is a paragraph of a document in our legal corpus. Instead of returning the full document, which is very long and consists of other less relevant information, \method\ returns the question $q$ in Table~\ref{tab:example} to the user. If the user finds that $q$ best paraphrases his concerns, the user will select $q$ and \method\ will bring to the user the answer $s_q$ as well as the document source, $d_q$. 
%Essentially, QBR plays the role of a domain expert. 
%Through \method, we provide free and easily accessible professional legal answers to the general public.

% Given a document collection that provides professional knowledge on law (such as the legal education website), \method\ first construct a question bank (QB) that contains frequently asked questions and their answers.
% A user simply need to verbally describe the legal situations she/he is facing, and \method\ would use the QB to help the user rephrase the question and return the relevant short answers.

\begin{table}[tbp]
\centering
\resizebox{.44\textwidth}{!}{
\begin{tabular}{p{11cm}}
\hline
\textbf{User Input:} I was out having a drink and I took a picture of a nightclub signboard. Out of nowhere, a man rushed up to me and started accusing me loudly, saying, ``Hey! What are you doing taking a picture of me!?'' He even demanded that I delete the photos. What is the relevant law in this situation? \\ %I felt threatened and so I did right away. \\
\hline
\hline
%\textbf{Expert-rephrased Question:} 
% \textbf{$q$}:
$\bm{q}$: What are the consequences if persons take photos in a public place without the photo subject's consent and cause any person to be reasonably concerned for his or her safety?\\
\hline
%\textbf{Expert Answer:} 
% \textbf{$s_q$}
$\bm{s_q}$: Persons who publish photos with captions that contain personal data of the photo subjects (without their consent) may have violated the Personal Data Privacy Ordinance. If such captions contain unjustified adverse comments on the photo subjects, the publishers may also have incurred civil liability for defamation. Furthermore, persons who take photos on such occasions (i.e. in a public place without the photo subject's consent) and cause any person to be reasonably concerned for his/her safety may have committed the offence of loitering or may be charged with behaving in a disorderly manner in a public place.\\
\hline
\end{tabular}
}
\caption{An example of user input and $\qspair{q}{s_q}$ pair}
\label{tab:example}
\end{table}

% The rest of the paper is organized as follows.
% We first describe \method\ and discusses the benefits a question bank 
% We then present experimental results and case studies on the impact of \method.
% Finally, we summarize the related works and conclude the paper.

Our key contributions are as follows:
\simplebullet\ We propose \method, a new question-bank-based approach to effectively answer legal questions by assisting users with fine-grained legal knowledge retrieval from a corpus of legal educational documents. We show that \method\ effectively simplifies the retrieval process, which is critical in community legal question-answering systems where users are typically non-legally-trained.
\simplebullet\ We explain the design of \method, particularly on how a question bank is effectively derived and employed to most accurately identify the matching knowledge units that best answer a given user's legal question. 
\simplebullet\ We deploy \method\ on an online public platform \footnote{\url{https://ai.hklii.hk/recommender/}}. We provide a case study to illustrate how such a platform helps the public solve their legal problems. This shows the social impacts of our platform.
\begin{comment}
\newline\noindent$\bullet$ We propose QBR, which employs a question bank and allows laypeople to query and receive legal information in plain language.
\newline\noindent$\bullet$ Through improving QBR's ability in identifying answer scopes, users can easily extract useful information simply by skimming through the returned question-answer pairs.
\newline\noindent$\bullet$ The QBR is available online to solve the increasing numbers of legal questions, which in turn promotes fair access to justice.
\end{comment}
\section{Question Bank and QBR}
\label{sec:qb}

In this section we give an overview of QBR highlighting four benefits of using a question bank.
\paragraph{Question Bank}
Let $D$ be a document collection that provides professional knowledge on a certain subject matter, where $d \in D$ denotes
a document that is comprised of paragraphs. 
Conceptually, each document provides information on a particular topic under the subject matter covering various aspects and details.
We consider each document $d$ to be composed of multiple 
{\it knowledge units}, each could be the subject of a user's enquiry.
For example, a document on a legal advice website that explains traffic offenses would consist of sections (knowledge units), each defines a specific type of offence and its penalty.
A Question Bank ($\qb$) is a collection of question-document-answer tuples $\qds$'s, where $q$ is a question whose answer
can be found in document (with id) $d_q$, and $s_q$ is an {\it answer scope} (or simply {\it scope}) that specifies the paragraphs within document $d_q$ 
that explicitly answer $q$. Each scope $s_q$ thus represents a single logical knowledge unit that answers a specific question ($q$).
We will elaborate on how $\qb$ is constructed from a document collection $D$ later in this section.
For the moment, we assume that the knowledge presented in the documents of $D$ is {\it well covered} by the $\qb$. 
That is, corresponding to each knowledge unit presented in $D$, there is one or more $\qds$ tuples in $\qb$ such that
the answer scope $s_q$ comprehensively conveys the knowledge.

For each document $d \in D$, we construct a {\it question set} $\qd{d}$ that includes all the questions in the question bank
whose answer scopes are found in $d$. Also, the corresponding answer scopes are collected into a {\it scope set} $\sd{d}$.
Formally,
\begin{scriptsize}
\begin{equation*}
\qd{d} = \{q | (q, d_q, s_q) \in \qb, d_q = d\};\;\; \sd{d} = \{s_q | (q, d_q, s_q) \in \qb, d_q = d\}.
\end{equation*}
\end{scriptsize}
Since each answer scope $s_q$ is considered a single logical unit of knowledge (that answers a specific question),
the scope set $\sd{d}$ specifies all the knowledge units found in $d$. Therefore, $\sd{d}$ 
provides a structured semantic representation of $d$ based on the knowledge it contains.
Moreover, the question set $\qd{d}$ gives a set of questions that are answered by the content of document $d$ and so they are highly 
relevant to $d$. 
This brings us to the first advantage of having a question bank:\\ 
{\bf Adv.\@ 1 (Document Augmentation)}: {\it A $\qb$ provides textual information $\qd{d}$ that describes each document $d$.}\\
Essentially, the representation of each document $d$ can be augmented to $d + \qd{d}$.
As we will see later,
this augmentation helps disambiguate 
documents, significantly improving retrieval accuracy.

\paragraph{\method} The \method\ method consists of two steps, namely, document retrieval followed by answer scope retrieval. 
Given a user input $u$, Step 1 locates the best matching documents $d^* \in D$. 
Then, given $d^*$, \method\ identifies the best matching scope $s^*$ within $d^*$ as the answer to the user query $u$.
Algorithm~\ref{alg:qbr} outlines the procedure. We elaborate on the two steps in the following descriptions.

\begin{algorithm}[tb]
    \caption{\method}
    \small
    \label{alg:qbr}
    \textbf{Input}: Question Bank $\qb$, User Input $u$, Embedding Function $T$, CL Adjusted Embedding $T'$\\
    \textbf{Output}: Document $d^*$, Scope $s^*$
    \begin{algorithmic}[1]
        \STATE  $//$ (Step 1) Document selection\label{alg:qbr:doc} 
        \STATE  $(\tilde{q}, d^*, \tilde{s}) \leftarrow
         \mathop{\arg \max}\limits_{(q, d, s)\in \qb} \frac{(T(u) \cdot T(q;s))}{(\lVert T(u) \rVert \cdot \lVert T(q;s) \rVert)}$;
        \vspace{10pt}
        \STATE  $//$ (Step 2) Scope disambiguation\label{alg:qbr:scope} 

        \STATE $S_{d^*} \leftarrow$ scope set of $d^*$
        \STATE $s^* \leftarrow
         \mathop{\arg \max}\limits_{s \in S_{d^*}} \frac{(T'(u) \cdot T'(s))}{(\lVert T'(u) \rVert \cdot \lVert T'(s) \rVert)}$;

        \STATE \textbf{return} $d^*, s^*$
    \end{algorithmic}
\end{algorithm}

\paragraph{Step 1: Document Retrieval} 
Traditional approaches to document retrieval measure a similarity $\simut{}{u}{d}$ between a user's input $u$ against each document $d \in D$
and then return the documents that give the highest similarity scores. 
There are many methods ranging from simple word matching (e.g., BM25) to lexical analysis and embedding techniques (e.g., BERT). 
These methods differ in the similarity functions they employ.
With the availability of a question bank, QBR matches user input ($u$) against {\it questions} ($q$'s) and {\it scopes} ($s$'s) that are found in the
$\qb$ instead of directly against the documents ($d$'s) themselves.
The rationale is twofold: 
First, user input is expressed in the form of a question and therefore matching $u$ against questions $q$'s are generally more effective compared with matching $u$ against documents $d$'s.
Secondly, a document $d$ could contain multiple knowledge units, some of which could be irrelevant to the user input. 
This extra (irrelevant) information in $d$ acts as noise, which lowers the effectiveness of the similarity measure.
Scopes ($s$'s), on the other hand, are fine-grained knowledge units. They provide more focused retrieval, which reduces the noise effect.

\begin{comment}
Therefore, besides enhancing document retrieval accuracy, another advantage of $\qb$ is:\\
{\bf Adv.\@ 2 (Fine-grained Retrieval)}: {\it A $\qb$ enables fine-grained scope-based retrieval.}\\
Note that scopes that match a user input address the user's enquiry directly. This saves the user's effort in extracting relevant details from matching documents.
\end{comment}
%\subsection{Domain-Specific Document Retrieval via QB}
%\label{sec:qb-retrieve}

\begin{comment}
In this section, we introduce how a QB helps document retrieval. 
Given a user input $u$, the document retrieval is done by first identifying the question and answer in QB that is most related to $u$. 
We then retrieve the document that is associated with the question and answer in the same QB entry.
The rationale is that the user input $u$ is in the form of a question, and matching $u$ with the questions (and answers) in the QB could make the retrieval more effective compared with matching $u$ with documents.
\end{comment}
\method\ employs a transformer neural network, denoted $T(x)$, that transforms a text sequence $x$ into a continuous vector representation in a semantically rich latent space.
Let $P$ be a collection of question-scope pairs (q-s pairs for short) derived from $\qb$, i.e., 
$P = \{(q, s) | (q, d, s) \in \qb\}$.
QBR identifies the q-s pairs that best ``match'' a user input $u$ by measuring the cosine similarity between each q-s pair and $u$. That is,
\begin{equation}
\label{eq:simut}
    \simut{}{u}{(q,s)} = (T(u) \cdot T(q;s))/(\lVert T(u) \rVert \cdot \lVert T(q;s) \rVert).
\end{equation}
Let $(q^*, s^*) = \argmax_{(q,s) \in P} \simut{}{u}{(q,s)}$ be the pair with the highest similarity score, and let $(q^*, d^*, s^*)$ be the corresponding entry in $\qb$. 
\method\ identifies $d^*$ as the best matching document.

\paragraph{Step 2: Scope Disambiguation with Contrastive Learning}
%\subsection{Contrastive Representation Learning for Scope Retrieval}
%\label{sec:contrastive}
Given the best matching document $d^*$, \method\ performs fine-grained knowledge retrieval by identifying 
the scope in document $d^*$ that best answers
a given user input $u$.
We remark that this scope-based retrieval is much more challenging than traditional document retrieval. 
The reason is that while different documents generally cover different topics, different scopes of the same document present various facts and details {\it of the same topic}. It is therefore much harder to discern the subtle differences among the knowledge units (scopes) within the same document to pinpoint the one that best
answers a user's enquiry.
Moreover, due to the wide knowledge gap between professional documents and novice users, a user input $u$ may not provide sufficient textual clues to 
disambiguate scopes. For example, a user concerned with ``defamation'' charges might actually mean ``libel'' or ``slander''.
The two cases could be individually explained in two different scopes, both referencing ``defamation''. 
%\ben{Might need a better example to replace the above one.}
%\ming{For example, Both ``{\it M1: At the court hearing (where the creditor is usually represented by a lawyer), the Court may grant a bankruptcy order against the debtor if...''}, and ``{\it M2: At the court hearing (the debtor can be represented by a lawyer), the court may grant a bankruptcy order against the debtor if...''} acknowledge the court's authority to grant bankruptcy order against the debtor under specific conditions. However, {\it M1} primarily centers on the creditor's viewpoint, and {\it M2} shifts the focus to the debtor's position.} % Page 98
\method\ improves scope-based retrieval by utilizing the QB and applying contrastive learning (CL) to modify the embedding function.
The idea is that each scope $s$ is associated with some questions in the QB, given by
\begin{equation}
    \label{eq:rs}
    R_s = \{q | \qds \in \qb, s_q = s\}.
\end{equation} 
These questions provide additional information to improve the separation of scopes' embedding based on the questions the scopes (knowledge units) answer.

Given a document $d$ and its scope set $S_d$, our objective is to apply contrastive learning (CL) to adjust the 
embedding function $T()$ such that the embedding vectors of the scopes in $d$ are sufficiently separated. 
The idea is to utilize the question set $Q_d$ and the scope set $S_d$ from $\qb$ to construct training examples.
Specifically, for each $s \in S_d$, we first obtain the set of questions $R_s$ that $s$ answers.
We then construct a positive example set ($\posex{s}$) and a negative example set ($\negex{s}$):
\begin{scriptsize}
\begin{equation}
\label{eq:posex-negex}
    \posex{s} = \{(q,s) | q \in R_s \};\;\; \negex{s} = \{ (q,s') | q \in R_s, s' \in S_d \setminus \{s\} \}.
\end{equation}
\end{scriptsize}
Note that for each $\qspair{q}{s}\in \posex{s}$, $s$ is an answer of $q$, while for each $\qspair{q}{s'} \in \negex{s}$, $s'$ is a scope
in document $d$ that does not answer $q$. 
Same as InfoNCE~\cite{infonce} with cosine similarity on normalized embedding~\cite{NT_xent}, we use softmax loss to differentiate positive examples from negative examples.
We fine-tune the embedding function $T()$ with the objective of pulling the embedding vectors $T(q), T(s)$ of positive examples $(q,s) \in \posex{s}$ closer and pushing those of negative examples in $\negex{s}$ farther.
We use the following equation for training loss:
%\begin{scriptsize}
    \begin{equation*}
    \resizebox{0.45\textwidth}{!}{$
        L_{\mathit{CL}}(s) = 
        -\sum\limits_{(q,s)\in \posex{s}}\log\frac{e^{\mathit{sim}(q,s)/\tau}}{e^{\mathit{sim}(q,s)/\tau}+\sum_{(q, s') \in \negex{s} }e^{\mathit{sim}(q, s')/\tau} } 
       $},
       \label{equ:loss}
    \end{equation*}
%\end{scriptsize}
%The 4th advantage $\qb$ brings is:\\
where $\tau$ is the temperature parameter and 
$\mathit{sim}(a,b)$ represents the cosine similarity of $T(a)$ and $T(b)$.
We use $T'()$ to represent the adjusted embedding function obtained by CL.
We remark that the purpose of $T'()$ is to provide an embedding that can better disambiguate the different scopes {\it within the same document}. 
On the other hand, the original function $T()$ is retained for the purpose of document retrieval.
With CL, 
\begin{comment}
the \method\ retrieval procedure is modified into a 2-step process.
Step 1 (doc. retrieval): the best matching document $d^*$ is evaluated and retreived using the original embedding function $T()$
(Equation~\ref{eq:simut}). Step 2 (scope identification): given the best-matching document $d^*$, 
\end{comment}
we measure the similarity
between the user input $u$ and the scopes in $S_d$ using the embedding function $T'()$ to identify the best-matching scope $s^*$ (see Algorithm~\ref{alg:qbr}). 
We also find an entry $\qspair{q}{s^*} \in \qb$ as a search result q-s pair to be displayed to the user (see Figure~\ref{fig:QBR}).
From this discussion, we see that the QB provides a means for scope-based knowledge retrieval.

{\bf Adv.\@ 2 (Fine-grained Retrieval)}: {\it A $\qb$ enables fine-grained scope-based retrieval. In particular, it provides training examples for contrastive learning for effective scope disambiguation.}\\
\vspace{-3pt}
%\paragraph{User Input (Story) Simulation}
\paragraph{GPT-Augmentation}
%As we discussed in the introduction and illustrated in Table~\ref{tab:example}, 
%a novice user's enquiry could be lengthy and noisy.
We augment the CL training data by employing ChatGPT~\cite{chatgpt} to generate user input that mimics novice users.
Specifically, for each document $d \in D$, we select 
two scopes in $S_d$ whose embedding vectors are the most similar.
These scopes are the hardest to disambiguate. For each such scope $s$, we consider any $\qspair{q}{s} \in \posex{s}$ 
and generate a few user inputs $\gptui{q}$ using ChatGPT with the prompt:
% ``{\it Given the following context, provide a realistic real-life scenario that a person who knows nothing about [DOMAIN] knowledge might encounter. Context: $q$.}''
% In the prompt, [DOMAIN] refers to the specific domain of interest, such as ``legal'' or ``medical''.
%\kevin{Removed [DOMAIN], changed to {\it legal}}
``{\it Given the following context, provide a realistic real-life scenario that a person who knows nothing about legal knowledge might encounter. Context: $q$.}''
%We remark that our approach can be applied in other domains by changing ``legal'' to the specific domain of interest in the prompt.
We then add $\qspair{\gptui{q}}{s}$ to $\posex{s}$ and $\qspair{\gptui{q}}{s'}$, where $s' \in S_d \setminus \{s\}$, to $\negex{s}$.

The \method\ procedure we have mentioned so far selects one matching document $d^*$ (Step 1) from which a qs-pair (Step 2) is obtained.
We extend the above procedure by retrieving the top-$k$ documents (and hence top-$k$ q-s pairs) if multiple ($k$) search results are desired. 
If so, the top-$k$ questions, say $q_1$, .., $q_k$ (and their respective answers $s_1$, .., $s_k$), are displayed to the user, who will then choose among the returned questions the ones that best match his/her enquiry.
Note that under QBR, this step of {\it result filtering} involves the user reading only the returned questions $q$'s, which are 
very short text.
%to identify the one that best expresses the user's enquiry. 
This is in sharp contrast to traditional document retrieval systems in which
a user needs to read through the much longer content of the returned documents to determine their relevancy.
Also, the relevant knowledge units are directly given by the answer scopes $s_i$'s. The user need not go through a complete document to locate the knowledge.
Furthermore, each returned question $q_i$ explains how its corresponding answer $s_i$ addresses the user's enquiry:
if $q_i$ is an accurate rephrase of the user's input, then $s_i$ is the desired answer
(see example in Table~\ref{tab:example}). This leads to the third advantage of a $\qb$:\\
{\bf Adv.\@ 3 (Explainability, Comprehensibility, and Efficiency)}: {\it Questions in a $\qb$ explain the relevancy of
retrieval results, helping users to efficiently comprehend the extracted knowledge.}

So far, we have assumed the availability of a QB. We end this section with a brief discussion on how a QB is obtained.
With advances in NLP techniques such as large language models (LLMs), there have been quite a few works on generating questions from text documents, especially in the area of Education, where
the interests lie in generating assessments (questions) to evaluate students' mastering of knowledge conveyed in course materials (documents).
% (See Section~\ref{sec:related} for references.)
For professional documents, \cite{ai_and_law_LQB} studies the problem of constructing a question bank from a corpus of 
web pages, each of which is a document that explains a specific legal topic in layperson's terms. 
In their study, {\it human-composed questions} (HCQs) and {\it machine-generated questions} (MGQs) are collected.
The HCQs are manually written by legal experts who were instructed to ask questions for every aspect covered by the documents and to identify the answer (scope) for the questions.
MGQs are machine-generated questions using the GPT-3 175B model. Readers are referred to~\cite{ai_and_law_LQB} for technical details.
In that study, the authors compare MGQs and HCQs w.r.t.\@ several
quantitative measures. Some advantages of MGQs over HCQs include lower cost and a higher number of questions. 
Moreover, it is reported that about 93\% of the documents' contents are covered by the MGQs.
Using machines to construct a QB is therefore a practical approach. 

\section{Experiments} % Here we only use legal dataset but in the above part we need to ignore the specific domain : legal retrieval
\label{sec:experiment}
In this section, we evaluate the performance of QBR. We describe the experiments conducted and present the results.
\begin{comment}
We conduct experiments and report the results in this section.
As our work improves the retrieval process of domain-specific documents, we conduct the experiments in the legal domain.
In Section~\ref{sec:exp-settings}, we describe the settings of the experiments.
We evaluate the effectiveness of the document retrieval via a question bank and discuss the results in Section~\ref{sec:exp-doclevel}.
In Section~\ref{sec:exp-contrastive}, we present the evaluation results of contrastive learning and demonstrate that our approach can significantly improve the accuracy in pinpointing the answer within a document.
\end{comment}

\subsection{Experiment Settings}
\label{sec:exp-settings}
% \kevin{I suggest changing this section to ``Dataset'', and then describe the experiments separately in 4.2 (document-level) and 4.3 (scope-level). The training details shall be moved to 4.3 where we actually do the training.}
% \ming{Keep it for now}
\paragraph{Data} 
The Community Legal Information Centre (CLIC)\footnote{\label{fn:clic}\url{https://clic.org.hk/en}} is a website that
provides legal information to the general public in Hong Kong. 
Legal knowledge under different topics is presented in thousands of web pages. 
In~\cite{ai_and_law_LQB}, human-composed questions (HCQs) and machine-generated questions (MGQs) are derived from these CLIC web pages.
For our experiment, we obtained a set of 1,359 CLIC pages as our document collection $D$.
Also, we obtained 15,333 HCQs and 23,238 MGQs from~\cite{ai_and_law_LQB} for a total of 38,571 questions (and their answer scopes)
as our question bank $\qb$. We use $\qbh$ and $\qbm$ to represent the sets of HCQs and MGQs, respectively. 
Note that 
our question bank 
$\qb = \qbh \cup \qbm$.

To perform CL training, we construct training examples using $QB$. 
We follow the procedure described in the previous section %Section~\ref{sec:qb} 
to generate positive and negative examples of 
q-s pairs
(Equations~\ref{eq:rs} and \ref{eq:posex-negex}).
The number of (+ve; -ve) examples obtained are (15,333; 125,977) and (23,238; 236,315) using $\qbh$ and $\qbm$, respectively.
%for a total of (38,571, 362,292) training examples.
We further augment the training set using GPT-augmentation.
For this step, we %apply scope selection to 
select two least-distinguishable scopes from each document.
Based on each selected scope $s$ and their associated questions $q$'s given in $\qb$, 
we synthesize user input $\hat{u}$'s to form positive and negative $\qspair{\hat{u}}{s}$ examples. %(see Section~\ref{sec:qb}).
We then sample (12,168; 72,747) examples from this pool and include those in the training set.
The final training set has (50,739; 435,039) examples in total.

Finally, we sample 1,000 positive $\qspair{\hat{u}}{s}$ examples that are not included in the training set as our test
set $U$. For each $\qspair{\hat{u}}{s} \in U$, the synthesized user input $\hat{u}$ has its ground-truth answer given by
the scope $s$, and the ground-truth document is the one that contains $s$.

\paragraph{Performance metrics}
Recall that \method\ carries out a two-step retrieval process, namely, {\it document retrieval} followed by {\it scope identification} within the retrieved documents. 
We therefore evaluate \method's performance based on its accuracy in these two steps.  
Consider a test case user input $u \in U$ whose answer scope $s^*$ is contained in document $d^*$.
For document retrieval accuracy, we consider the top-k
documents retrieved by \method\ from dataset $D$. The {\it recall@k} of the retrieval result is 1 if
$d^*$ is among the top-k documents; 0 otherwise. 
Also, the {\it reciprocal rank} ($\mathit{rr}$) of the result is $1/i$, where $i$ is the rank of $d^*$ in the result. (Rank = 1 if $d^*$ is ranked first in the result list; $\mathit{rr}$ = 0 if $d^*$ is out of top-k.)
We report the average recall@k and mean reciprocal rank (denoted by $\mrrd$) computed over the whole test set $U$ of 1,000 cases.
%A higher recall@k implies better performance because it is more likely that the target document is shortlisted and displayed to
%the user for a successful retrieval.
%A lower MRR is better performance because the target document is ranked higher in the returned list.

For scope identification accuracy, we assume the target document $d^*$ is successfully retrieved, 
and so we consider the order in which \method\ ranks the scopes within the document (i.e., the scopes given in the scope set $S_{d^*}$). 
We report the average reciprocal rank of target scopes $s^*$ of all test cases in $U$.
We denote this measure $\mrrs$ to distinguish it from the ranking measure $\mrrd$, which was defined for the document retrieval step.
We also report the average accuracy ($\acc$), defined as the fraction of cases in which \method\ pinpoints the 
correct scope $s^*$ as the answer to the user input (i.e., $s^*$ identified as top-ranked among all scopes in $d^*$).
\begin{comment}
In the experiments, we evaluate the performance of the retrieval models from two dimensions. 
1) {\bf Accuracy}: Retrieving the accurate document, or retrieving the correct scope at a more granular level and presenting it to the user, is critical. 
We measure the accuracy by recall@k. 
For each input, since there is only one ground truth document/scope to retrieve, recall@k would be 1 if the ground truth ranked within top-k, and 0 otherwise.
We report the recall@k averaged over the entire test set.
2) {\bf Ranking}: It is also crucial to prioritize and present highly relevant documents to users. We used the mean reciprocal rank (MRR) as an evaluation measurement. 
Specifically, we only assess the top-5 documents/scopes retrieved with a user story, and compute the reciprocal rank.
If the ground truth is not ranked within top-5, we assign 0 as the reciprocal rank.
\end{comment}

\paragraph{Baseline Methods} %\kevin{shortened}
%\subsection{Baseline Methods}
We compare \method\ against a wide range of retrieval methods. These methods differ in the way they measure similarity between
user input (queries) against documents in a corpus.
In particular, we consider three categories of approaches.
{\bf Lexical} models (BM25~\cite{bm25}) measure similarity based on 
word occurrences. 
{\bf Sparse} retrieval models, (SPARTA~\cite{sparta}, docT5query~\cite{docTTTTTquery}), encode terms' 
occurrences in documents and queries using high-dimensional, sparse vectors.
{\bf Dense} retrieval models (BERT~\cite{Sentence-BERT}, TinyBERT~\cite{Sentence-BERT}, TAS-b~\cite{tas}, RoBERTa~\cite{roberta}, DPR~\cite{DPR}, MPNet~\cite{mpnet}) %, on the other hand, 
utilize low-dimensional, dense vector representations for documents and queries. 

\begin{comment}
Specifically, BERT~\cite{Sentence-BERT} is a BERT-based model that is trained on the MSMARCO dataset for semantic passage retrieval. TinyBERT~\cite{Sentence-BERT}, a variant of BERT, offers a more compact model, sharing similarities with TAS-b~\cite{tas}, with BERT being approximately 1.7 times larger than both. 
RoBERTa~\cite{roberta} is a robustly optimized version of BERT. ANCE~\cite{ance} is a variant of RoBERTa. DPR~\cite{DPR} utilizes the BERT architecture but is tailored for question answering tasks. 
MPNet~\cite{mpnet} performs fine-tuning for sentence embedding. 
SBERT
%, deriving from the MiniLM~\cite{minilm} model, 
is a lightweight and distilled version of BERT.
%, is fine-tuned on the MSMARCO dataset for passage retrieval.
\end{comment}

\method\ employs an embedding function $T()$. In our experiment, \method\ uses MPNet's embedding (as function $T()$) as the default.
We remark that although \method\ uses MPNet's embedding function, it differs from the original MPNet method because \method\ uses the
QB to enhance retrieval performance.
First, in document retrieval, \method\ compares the embedding vectors of user input $u$ against the q-s pairs in $\qb$ (Equation~\ref{eq:simut}), instead of against the document embedding vectors. 
Secondly, in scope identification, \method\ uses CL to revise the embedding function (to $T'()$) to better disambiguate scopes within the same document. 
The details of CL model training are reported in the technical appendix\footref{fn:appendix}. %\kevin{appendix [CL Model Training]}

\begin{table*}[t]
\centering
\resizebox{.75\textwidth}{!}{
\centering
\begin{tblr}{
  cells = {c},
  cell{1}{1} = {r=2}{},
  cell{1}{3} = {c=2}{},
  cell{1}{5} = {c=8}{},
  % cell{3}{1} = {r=4}{},
  % cell{7}{1} = {r=4}{},
  vline{2-3,5,13} = {-}{dotted},
  hline{1,3,7} = {-}{},
  %hline{7} = {-}{},
  hline{6} = {-}{dashed},
}
               & Lexical & Sparse Models &            & Dense Models   &        &         &        &        &        &        &         &  \\
& BM25    & SPARTA & docT5query & TinyBERT & BERT   & RoBERTa & ANCE   & DPR    & TAS-B  & SBERT  & MPNet   & QBR\\
Recall@1 & 0.2540  & 0.1890 & 0.3300     & 0.2770   & 0.3840 & 0.3300  & 0.3670 & 0.1920 & 0.3900 & 0.4520 & 0.4500     & {\bf 0.5400}\\
Recall@3    & 0.4160  & 0.3170 & 0.4950     & 0.4130   & 0.5250 & 0.5100  & 0.5540 & 0.3150 & 0.5700 & 0.6370 & 0.6670     & {\bf 0.7230}\\
Recall@5    & 0.5090  & 0.3820 & 0.5800     & 0.4940   & 0.6060 & 0.5750  & 0.6250 & 0.3850 & 0.6520 & 0.7180 & 0.7360    & {\bf 0.8050}\\
$\mrrd$         & 0.3584  & 0.2697 & 0.4333     & 0.3666   & 0.4763 & 0.4364  & 0.4801 & 0.2758 & 0.4993 & 0.5639 & 0.5739     & {\bf 0.6482}
\end{tblr}
}
\caption{Document retrieval performance}
\label{tab:document_retrieval_result}
\end{table*}

\begin{table*} [t]
    \centering
    \resizebox{0.64\textwidth}{!}{
    \begin{tblr}{cells={c},
    vline{2,10,11} = {-}{dotted},
    hline{3} = {-}{dashed},
    }
    \hline
        ~           & TinyBERT & BERT   & RoBERTa & ANCE   & DPR    & TAS-B  & SBERT  & MPNet  & QBR  & $\qbrnogpt$ \\
    \hline
        $\acc$ & 0.4390   & 0.4650 & 0.4500  & 0.4410 & 0.3580 & 0.4390 & 0.4870 & 0.5000  & {\bf 0.8370} & 0.6460          \\
        % Recall@3    & 0.8338   & 0.8402 & 0.8333  & 0.8133 & 0.7608 & 0.8287 & 0.8578 & 0.8675 & 0.9578 & 0.8758         \\
        % Recall@5    & 0.9341   & 0.9339 & 0.9380  & 0.9237 & 0.9005 & 0.9309 & 0.9507 & 0.9495 & 0.9819 & 0.9471          \\ \cline[dashed]{2-12}
        $\mrrs$         & 0.6638   & 0.6763 & 0.6683  & 0.6570 & 0.5964 & 0.6631 & 0.6963 & 0.7061 & {\bf 0.9100} & 0.7860          \\
    \hline
    \end{tblr}
    }
    \caption{Scope identification performance}
    \label{tab:scope_retrieval_result}
\end{table*}

% \begin{table*} [htbp]
%     \resizebox{\textwidth}{!}{
%     \centering
%     \begin{tblr}{ccccccccc|c|cc}
%     \hline
%         ~           & TinyBERT & BERT   & RoBERTa & ANCE   & DPR    & TAS-B  & SBERT  & MPNet  & QBR   & w/o Story & w/o Scope Selection  \\
%     \hline
%         Recall@1 & 0.4390   & 0.4650 & 0.4500  & 0.4410 & 0.3580 & 0.4390 & 0.4870 & 0.5000  & 0.8370 & 0.6460    & 0.7730        \\
%         Recall@3    & 0.8338   & 0.8402 & 0.8333  & 0.8133 & 0.7608 & 0.8287 & 0.8578 & 0.8675 & 0.9578 & 0.8758    & 0.9182        \\
%         Recall@5    & 0.9341   & 0.9339 & 0.9380  & 0.9237 & 0.9005 & 0.9309 & 0.9507 & 0.9495 & 0.9819 & 0.9471    & 0.9619        \\ \cline[dashed]{2-12}
%         MRR         & 0.6638   & 0.6763 & 0.6683  & 0.6570 & 0.5964 & 0.6631 & 0.6963 & 0.7061 & 0.9100 & 0.7860    & 0.8665        \\
%     \hline
%     \end{tblr}
%     }
%     \caption{Results of scope retrieval within document with and without contrastive learning}
%     \label{tab:CL_overall}
% \end{table*}

%\input{0c_table_3.tex}
\vspace{-3pt}
\subsection{Results}
\vspace{-2pt}
\paragraph{Document Retrieval}
Table~\ref{tab:document_retrieval_result} shows document retrieval performance comparing \method\ against 11 baseline methods.
Among the baselines, the lexical model BM25, which could be viewed as a robust benchmark for generalization~\cite{beir}, 
performs better than SPARTA (sparse model) and DPR (dense model) in terms of both recalls and $\mrrd$, while having comparable performance to TinyBERT. 
Among the sparse models, docT5query employs document expansion to capture an out-of-domain keyword vocabulary, which enhances performance over
SPARTA. 
%enhancing the accuracy of document retrieval. 
% However, sparse models employing term weighting methods, such as SPARTA, faced limitations. 
Dense models, particularly SBERT and MPNet, exhibit superior performance, showcasing a proficiency in understanding contextual information. 
%Nevertheless, the effectiveness of the re-rank method is hindered by its reliance on the outcomes of the initial stage, resulting in suboptimal performance.\kevin{Worse than BM25?}
With the exception of DPR and the small TinyBERT model, dense models, particularly MPNet, are better performers.
\method\ uses MPNet for its base embedding function, but it applies the QB in document retrieval by matching user input against
the QB entries. Table~\ref{tab:document_retrieval_result} shows that this approach significantly improves performance.
For example, for Recall@1, \method\ (0.5400) is more than twice better than the traditional BM25 method (0.2540) and is much better than
the best of the baselines, MPNet (0.4500).
%In Section~\ref{sec:qb}
Previously we mentioned 3 advantages of using a QB. 
The results in Table~\ref{tab:document_retrieval_result} show strong evidence of Adv. 1 ({\it Document Augmentation});
Questions provided in the QB serve as important information that facilitates document representation, leading to much more effective document retrieval.

We remark that traditional retrieval methods, such as the baselines shown, return the highly ranked {\it documents} to users in response to their enquiries.
To filter the returned results, a user needs to read through the returned documents to find the answer. 
In contrast, \method\ returns $\qspair{q}{s}$ pairs (see Figure~\ref{fig:QBR}) and the user needs to
read only the questions $q$'s returned and (if needed) their answer scopes $s$'s to understand if any of the returned entries answer the enquiry
(see Table~\ref{tab:example} for an example).
The amount of content read is thus much smaller. 
For example, an average document in our collection $D$ contains 5 scopes (i.e., 5 knowledge units).
If we use MPNet as the method for document retrieval, then the probability of finding the answer from the top-ranked document
is 45\% (MPNet's Recall@1 = 0.4500). For the same amount of time (to read one document with 5 scopes), with \method, the user could have browsed 
5 $\qspair{q}{s}$ entries in the returned results. The probability of the user finding the answer among them is 80.5\% (\method's Recall@5 = 0.8050),
which is almost twice as likely as for the case of MPNet. 
From the perspective of user search efforts, \method\ is much more efficient than all other approaches.
The experimental results thus show evidence of Adv. 3 ({\it Explanability, Comprehensibility, and Efficiency}).

\paragraph{Scope Identification}
Step 2 of \method\ involves identifying the correct scope $s^*$ within the target document $d^*$.
This step can also be done using the baseline methods by applying them to rank the scopes of $d^*$. 
However, \method\ utilizes the QB to perform contrastive learning (CL) to adjust its embedding function ($T'()$) so as to 
better disambiguate scopes within the same document.
Furthermore, \method\ uses GPT to augment the CL training set. 
Given the correct document $d^*$, Table~\ref{tab:scope_retrieval_result} shows scope identification performance comparing \method\ against 
the dense model baselines. 
The column labeled ``$\qbrnogpt$'' refers to \method\ without GPT-augmentation in CL training. 
From the table, we see that none of the baselines has its $\acc$ exceed 0.5. 
That means they select the wrong answer scope more often than they pick the right one.
Scope identification is therefore a very difficult task due to the very similar semantics of the scopes within the same document.
By applying CL to obtain a more discerning embedding function $T'()$, \method\ gives a much higher $\acc$ at 0.837.
Moreover, \method\ has an $\mrrs$ of 0.91, which is very close to 1.0. This indicates that even for the cases
where \method\ does not rank $s^*$ first, $s^*$ is ranked very highly by \method. 
By comparing the performance of \method\ with $\qbrnogpt$, we see a significant drop in $\acc$ (0.8370 $\rightarrow$ 0.6460) 
and $\mrrs$ (0.9100 $\rightarrow$ 0.7860) if we take
away GPT-augmentation. This shows that the training examples obtained via GPT to mimic user input is highly effective.
Nevertheless, the scores of $\qbrnogpt$ are still way higher than those of the baselines. This again shows the effectiveness of CL.
These results support Adv. 2 (Fine-grained Retrieval).
%\newline

%\kevin{Appendix: 0c\_table\_3, 0c\_table\_4, QB quality, QB size, medical}

\vspace{-3pt}

\subsection{Demo and Additional Experiments}
\vspace{-2pt}
We have conducted additional experiments to further assess \method’s effectiveness. Additionally, we have deployed QBR on a real platform to help the public comprehend the law based on their specific legal circumstances. Due to space constraints, we provide a concise summary of our experimental findings. For more detailed information about the experimental results and a demonstration of the deployed platform, please refer to the technical appendix\footref{fn:appendix}.
%We have conducted additional experiments to further evaluate \method. Moreover, 
%we have deployed \method\ on a real platform to assist the public in understanding the law based on their legal situations.  
%Due to space limitations, we give a brief summary of our experimental findings. 
%Further details of the experimental results and a demo of the deployed platform are given in the technical appendix\footref{fn:appendix}. 
\paragraph{QB Quality}
We investigate how the QB quality affects \method's performance.
Intuitively, a good QB should (1) have rich contents (i.e., enough questions) that cover all knowledge units presented in
the documents and (2) be well-phrased and relevant to the knowledge units~\cite{ai_and_law_LQB}.
First, we investigated the performance of \method\ w.r.t. the size of QB.
We observe that \method's performance progressively improves as we increase the QB's size. 
However, we observe that even a small QB drastically improves scope identification accuracy ($acc$).
Specifically, a (small) QB with 10K questions achieves an $acc$ of 0.719, which is much higher than
MPNet ($acc$ = 0.500).
This shows that the QB provides critical information for disambiguating scopes and our CL approach
is highly effective even with a small QB. 
We further evaluate \method\ using three versions of the question bank: $\qbh$, $\qbm$, and $\qb = \qbh \cup \qbm$.
We observe that $\qbh$ and $\qbm$ give very similar performance with $\qbh$ having a slight edge
over $\qbm$. Also, both of them outperform MPNet by significant margins.
The complete $\qb$ gives the best performance, showing the complementary nature of machine and human questions.

\vspace{-3.5pt}
\paragraph{Language Models} We conducted experiments using different language models (in addition to MPNet) to derive the embedding
function $T()$ \method\ employs. We observe similar performance advantages of \method\ over other methods. 
\method\ is therefore a general approach that can work with different representation techniques.

\vspace{-3.5pt}
\paragraph{Scalability} Our system is efficient and scalable. Query execution time is dominated by the step that identifies the most relevant $\qspair{q}{s}$ to a user query. This step can be efficiently processed using a vector database. For example, with a QB of 38,571 questions, the average search time is 0.018s. We have conducted an experiment changing the QB size and found that the search time stays fairly stable. The scalability comes from effective vector indexing.
\vspace{-3.5pt}
\paragraph{Generalization to Other Domains} 
In addition to legal knowledge retrieval, our \method\ approach can be applied to various professional domains, including medical and financial. To illustrate \method’s versatility, we extend our study to medical knowledge retrieval. We utilized \method\ with medical data and conducted comparable experiments.
Similar conclusions regarding the performance of \method\ on document-level and scope-level retrieval can be drawn, which underscores the wide applicability of this method across different domains.

\section{Social Impact and Case Studies}

We have deployed QBR on the online legal information
platform CLIC\footref{fn:clic} 
and conducted user studies. 
Consent to participate in the studies have been duly obtained from all participants.
In the studies, participants were presented with hypothetical scenarios and were instructed to conduct searches for legal information using both their customary approach and QBR. Subsequently, they were asked to compare their search results. Some of the participants were NGO workers who provide services to families and victims of domestic violence cases.
All participants provided overwhelmingly positive feedback regarding QBR’s efficacy. For instance, in one scenario, participants were tasked with providing guidance to a domestic violence victim on divorce procedures and available protective measures for them and their children. Although participants were aware of the existence of court orders, they could not recall the precise legal action appropriate for the case. Many of them searched for ``protection order'' when the term they needed to find was ``injunction order''. 
%Despite being unable to locate the relevant content using their traditional search methods, 
They were able to find pertinent and helpful information on the platform with QBR
but their searches failed when they follow their traditional keyword search methods.
%that would have otherwise been overlooked with QBR. 
They all perceived QBR to be highly advantageous to their work since even experienced social workers may be perplexed by complex legal terminology. 
We also study how the information on CLIC and QBR can be utilized by a social work NGO for consultations on youth matters. 
The social workers believe that legal information written in layperson's terms can help them ``thoroughly understand the law in order to effectively assist clients and their parents'' when they come into contact with individuals who have been arrested.
\begin{comment}
%Since the beta launch of the QBR, a total of 
Our platform received 837 user searches in a 57-day period. Log data analysis indicates that members of the public have utilized the tool to seek answers to various aspects of daily life. Notably, users often provided scenario descriptions without including the relevant keywords for the legal issues, yet the system performed well and was able to identify the issues in question. 
This success is mostly attributable to how \method\ utilizes the question bank in paraphrasing user queries (see example in Table~\ref{tab:example}.)
\end{comment}
The QBR tool has garnered positive feedback and has proven to be effective in aiding members of the public in resolving legal problems. 

Using \method\ on a legal information platform like CLIC is impactful in upholding social justice, as the tool 
empowers the underprivileged community to be better legally informed.  
As an example, we interviewed a CLIC user, who works in the field of accounting and financial planning.
The user faced two litigation cases: one involves himself and the other involves his small company.
As the plaintiff had engaged a team of legal practitioners, the interviewee faced a knowledge imbalance.
He tried to seek help from various channels such as government legal advice schemes and court guidance notes to no avail. 
%but regrettably those were applicable only to individuals, not small companies.
He felt overwhelmed as he did not understand the legal procedures due to the high technicality and brevity of information.
%While attempting to refer to the guidance notes provided by the court online, they only provided a brief overview and did not clarify things for him.
Eventually, he tried CLIC and was able to comprehend the law and the relevant procedures with the information the CLIC platform provided.
Without hiring a legal representative, he was able to 
defend his cases and deter the opposing parties from pursuing the case further.
As he said in the interview, ``CLIC has been extremely helpful to us in preparing for the litigation''.
\vspace{-2pt}
\section{Related Works}
\label{sec:related}
%\miniheading{Domain-Specific Information Retrieval}
\paragraph{Comprehensibility of Legal Information}
\cite{mommers2011access} categorizes the accessibility of legal information into three levels: primary availability, where documents are available and searchable online; secondary availability, where links are established between relevant documents; tertiary availability, where contents are clarified and translated into languages understandable by the target audience.
\cite{mommers2011access} further studies two legal database websites in Europe and finds that tertiary availability is largely ignored.
\cite{dyson2017access} conducts a readability study on 407 passages extracted from Legal Services Corporation-sponsored websites and find that most of them are beyond comprehension by normal citizens, which contradicts the aim of the legal aid to serve those with low income and low literacy.
\cite{curtotti2013right} analyzes the language features and study the readability of Australian legal documents.
Aligned with previous empirical studies~\cite{pi2000report,tanner2002seventeen,abrahams2003efficacy}, they conclude that the linguistic characteristics of legal documents are quite different from normal English and that legal articles are generally more difficult to read.

\paragraph{Domain-Specific IR}
% domain-specific retrieval such as legal retreival
% fine-grained retreival 
%relevant as much as possible
\cite{InPars} underscores the significance of employing domain-specific training data over generic datasets. Notably, the study introduces a novel technique that leverages LLMs for the generation of synthetic training data, specifically tailored to IR tasks. 
%Expanding the domain-specific paradigm,
\cite{LER} introduces the task of Legal Evidence Retrieval (LER) with the objective of advancing real-world applications in Legal AI. The primary goal is to facilitate judges in efficiently locating pertinent oral evidence related to a given fact. However, a common limitation across these approaches is the potential oversight of user-centric considerations. Non-specialized users may lack familiarity with domain-specific terms and concepts, whichs %could
impede their ability to formulate accurate and relevant search queries. 

\paragraph{Contrastive Learning (CL)}
Recent research has explored the use of CL techniques to enhance retrieval systems. Ma, et al. \cite{ma2021contrastive} introduces a novel method for fine-tuning neural rankers, enhancing their robustness to out-of-domain data and query perturbations. \cite{RIR} explores self-supervised CL methods in Reviewed-Item Retrieval (RIR) task. \cite{robust_cl_typos} utilizes all available positive examples and implement multi-positive contrastive learning to address the issue of queries with typos, thereby enhancing the robustness of dense retrievers.
%proposes a supervised CL for fine-tuning pre-trained language models, addressing the limitations of cross-entropy loss in few-shot learning. 
Drawing insights from these works, our research endeavors to combine QB with CL, aiming to refine information retrieval at a finer granular level and to assist novice users.

%\vspace{-5pt}
\vspace{-2pt}
\section{Conclusion}
\label{sec:conclusion}

In this paper we proposed \method\ to perform fine-grained domain-specific knowledge retrieval for non-expert users. 
The challenge involves managing noisy and imprecise user input, as well as distinguishing and ranking semantically closely related scopes within the same document. 
\method\ tackles the problems through the use of a question bank (QB), which provides three advantages.  
By matching user input against QB's entries, and performing contrastive learning based on QB's data,
we show that \method\ significantly outperforms existing methods in terms of both document retrieval and scope identification.
Moreover, by returning re-phrased questions and answer scopes instead of documents, \method\ makes
retrieval results more comprehensible and explainable, leading to a more efficient user experience. 
In case studies, we showed that \method\ makes a social impact by helping users resolve day-to-day legal issues.

%\kevin{Appendix: Demo (Input) I am in an abusive relationship with my husband. I am considering divorce, but I am not familiar with the procedures and cost involved. Also, I am afraid that seeking a divorce might provoke revenge from my husband. I am unsure about what steps to take and how to ensure my daughter's and my safety.}

%Through extensive experiments, we systematically examined and analyzed the effectiveness of \method. In the future, we plan to investigate more possible applications of the question bank including conversational retrieval. It involves proactively posing questions to users, thereby further advancing the capabilities and user-centric nature of our proposed solution.
%The experimental results demonstrate..

\section*{Acknowledgement}

This research is supported by the WYNG Foundation (AR25AG100407).

\bibliographystyle{named}
\bibliography{ijcai25}

\end{document}